\documentclass[conference]{IEEEtran}
\IEEEoverridecommandlockouts
\usepackage{cite}
\usepackage{amsmath,amssymb,amsfonts}
\usepackage{algorithmic}
\usepackage{graphicx}
\usepackage{booktabs,tabularx,array}
\usepackage{textcomp}
\usepackage{amssymb}
\usepackage{xcolor}
\usepackage{tikz}
\usepackage[hidelinks]{hyperref}
\usetikzlibrary{positioning,calc,matrix,arrows.meta}
\def\BibTeX{{\rm B\kern-.05em{\sc i\kern-.025em b}\kern-.08em
    T\kern-.1667em\lower.7ex\hbox{E}\kern-.125emX}}
\begin{document}

\title{Multi-Speaker Conversational Audio Deepfake: Taxonomy, Dataset and Pilot Study\\
\thanks{}
}

\author{\IEEEauthorblockN{Alabi Ahmed}
\IEEEauthorblockA{\textit{Univ. of Maryland, Baltimore County} \\
United States \\
aahmed17@umbc.edu}
\and
\IEEEauthorblockN{Vandana Janeja}
\IEEEauthorblockA{\textit{Univ. of Maryland, Baltimore County} \\
United States \\
vjaneja@umbc.edu}
\and
\IEEEauthorblockN{Sanjay Purushotham}
\IEEEauthorblockA{\textit{Univ. of Maryland, Baltimore County} \\
United States \\
psanjay@umbc.edu}
}

\maketitle

\begin{abstract}
The rapid advances in text-to-speech (TTS) technologies have made audio deepfakes increasingly realistic and accessible, raising significant security and trust concerns. While existing research has largely focused on detecting single-speaker audio deepfakes, real-world malicious applications with multi-speaker conversational settings is also emerging as a major underexplored threat. To address this gap, we propose a conceptual taxonomy of multi-speaker conversational audio deepfakes, distinguishing between partial manipulations (one or multiple speakers altered) and full manipulations (entire conversations synthesized). As a first step, we introduce a new Multi-speaker Conversational Audio Deepfakes Dataset (MsCADD) of 2,830 audio clips containing real and fully synthetic two-speaker conversations, generated using VITS and SoundStorm-based NotebookLM models to simulate natural dialogue with variations in speaker gender, and conversational spontaneity. MsCADD is limited to text-to-speech (TTS) types of deepfake. We benchmark three neural baseline models; LFCC-LCNN, RawNet2, and Wav2Vec 2.0 on this dataset and report performance in terms of F1 score, accuracy, true positive rate (TPR), and true negative rate (TNR). Results show that these baseline models provided a useful benchmark, however, the results also highlight that there is a significant gap in multi-speaker deepfake research in reliably detecting synthetic voices under varied conversational dynamics. Our dataset and benchmarks provide a foundation for future research on deepfake detection in conversational scenarios, which is a highly underexplored area of research but also a major area of threat to trustworthy information in audio settings. The MsCADD dataset is publicly available\footnote{\url{https://github.com/MultiDataLab/MultispeakerDeepfakes}}
to support reproducibility and benchmarking by the research community.
\end{abstract}

\begin{IEEEkeywords}
Audio deepfake, Artificial Intelligence, Multi-Speaker, Conversational Deepfakes, Text-to-speech Systems, Generative AI
\end{IEEEkeywords}

\section{Introduction}

The advent of artificial intelligence (AI) has enabled the creation of highly convincing audio deepfakes, which have become a growing concern due to their potential misuse \cite{smith2021, harwell2021, stupp2022} in various sectors, including media, telecommunications, and cybersecurity. Audio deepfakes are synthetic audio recordings generated by AI techniques \cite{wang2018, sisman2020,shen2018,jin2017,kim2021, wang2017, wang2021, tan2021,popov2021} such as text-to-speech (TTS) synthesis, which can be indistinguishable, or nearly so, from real human speech. These deepfakes can be used to impersonate individuals, manipulate messages, or deceive audiences in ways that were previously unimaginable \cite{khanjani2023, almutairi2022,todisco2019}.

Audio deepfake detection refers to the task of identifying speech that has been synthetically generated or manipulated using artificial intelligence techniques to closely mimic real human speech in a deceptive manner\cite{khanjani2023, almutairi2022}. While research on detecting audio deepfakes \cite{todisco2019b,liu2021,khanjani2023b} has made significant strides, most efforts have concentrated on detecting deepfakes in single-speaker, or single-stream audio scenarios. These studies have typically used datasets such as ASVspoof \cite{wang2020, delgado2021} and ADD 2022/2023 \cite{yi2022b,yi2023}, which focus on identifying fake audio produced by individual speakers, either synthesized from text or altered through voice conversion.

\textit{\textbf{Motivation:} In September 2023, just two days before Slovakia’s parliamentary elections, a fabricated audio recording surfaced online. Wired and CNN \cite{wired-slovakia-deepfakes, cnn-election-deepfake-threats} reported that the audio clip simulated a conversation between opposition leader Michal Šimečka and journalist Monika Tódová, in which they allegedly discussed vote-rigging and ballot purchases. Both individuals quickly denounced the audio clip as fake, and subsequent fact-checking confirmed the presence of AI-generated manipulation. However, the recording was released during the legally mandated pre-election silence period, limiting opportunities for candidates or media to rebut the disinformation. The incident spread widely on social platforms and may have influenced public perception in a tightly contested election. This situation highlights the unique threat of multi-speaker conversational audio deepfakes in real-world, and demonstrates the need of further research on Multi-speaker conversational audio deepfake content.}

As demonstrated by this motivating scenario, subtle nuances in human speech, when carefully analyzed can provide critical cues for identifying manipulated audio. This observation underscores the importance of advancing research into the diverse manipulation scopes that can emerge in multi-speaker conversational audio deepfakes, where conversational dynamics introduce additional layers of complexity. Multi-speaker coversational deepfakes may emerge in settings, such as 1-1 conversations, interviews, debates, where more than one individual is speaking in the same audio clip. This is different from a scenario where a dataset may have multiple clips with mutliple speakers who may not be in the same audio clip (this is seen in the MADD \cite{qi2024} dataset). Detecting deepfake audios in multi-speaker conversational environments introduces new challenges due to factors such as speech overlap, speaker diarization, background noise and acoustic disturbance. To the best of our knowledge, no prior research has specifically focused on multi-speaker audio deepfake detection in conversational settings with multiple speakers in same audio clips, and existing datasets are ill-suited for training models in such environments. The need for datasets that specifically focus on such multi-speaker audio deepfakes is critical for advancing detection techniques and ensuring that these models can handle the growing risks associated with AI-generated fake content in conversational settings.
This gap in research is critical because deepfakes are not only being used in isolated speech scenarios but can infiltrate conversational contexts, such as corporate meetings, public debates, conferences and media interviews, where multiple individuals may be involved in dynamic dialogues. The risks posed by deepfakes in these settings are amplified, especially in instances of telecommunication fraud, political manipulation, and corporate espionage.

\textbf{Contribution:} Our key contributions are as follows:
\begin{itemize}
    \item We propose a conceptual taxonomy of multi-speaker conversational audio deepfakes, distinguishing partial manipulations (affecting one or multiple speakers) from full deepfakes (all speakers synthetic)
    \item We create  a new Multi-speaker Conversational Audio Deepfake Dataset (MsCADD): 2,830 audio clips of naturalistic conversations (2 speakers, mixed genders, diverse content) with clean and noisy backgrounds.
    \item We benchmark three baseline detection methods on this dataset. Each is evaluated with accuracy, true positive rate, true negative rate, and F1 score.
\end{itemize}

\textbf{In this paper we limit the scope to text-to-speech (TTS) types of deepfakes to demonstrate the need for multi-speaker detection and the need for new approaches. }

The rest of the paper is organized as follows. Section II discusses the background and the related work. In Section III we discuss multi-speaker deepfake taxonomy and dataset. Section IV discusses the experimental results of the baseline detection methods and in Section V, we present the conclusion and future work.

\section{Background and Related Work}

\subsection{Overview of Audio Deepfake Detection}

There are two primary categories of approaches for detecting spoofed and deepfake audio: Machine Learning (ML) and Deep Learning (DL) \cite{almutairi2022}. Classical ML models such as Random Forest, Logistic regression, Support Vector Machine (SVM), and K-Nearest Neighbors (KNN) have generally produced limited performance, with reported accuracies reaching only about 0.67 \cite{khochare2021deep, khanjani2023b}  on the FoR dataset \cite{reimao2019}.

Deep learning methods leverage neural network architectures that are better suited for modeling complex audio patterns. As deep learning technologies, especially generative models like Generative Adversarial Networks (GANs), have improved over time, the ability to create lifelike audio has become easier and more accessible \cite{khanjani2023}. Early detection methods focused on the detection of synthesized speech created through text-to-speech (TTS) systems. Such approaches have been successful in identifying deepfakes in controlled environments. ResNet-based architectures have been employed for audio deepfake detection \cite{chen2017resnet} and subsequently enhanced to improve performance and address generalization challenges \cite{chen2020generalization}. However, ResNet models remain computationally intensive and often struggle to adapt to unseen data. Other studies have explored Temporal Convolutional Networks (TCNs) on deepfake audios \cite{khochare2021deep}. The ASVspoof 2021 Challenge \cite{delgado2021} provided a standardized benchmark by proposing four baselines across three tasks.

Models trained to detect single-speaker deepfake audios, especially those generated by advanced TTS models like Tacotron \cite{wang2017}] or WaveNet \cite{oord2016}, often rely on spectral features \cite{khanjani2023}  such as Mel-frequency cepstral coefficients (MFCCs), spectrograms, and raw audio waveforms to distinguish between genuine and synthetic speech.

\subsection{Existing Datasets in Deepfake Audio Detection}

Current research on audio deepfake detection has primarily focused on single-speaker datasets, many of which contain utterances generated with text-to-speech (TTS) techniques. Table~\ref{tab:dataset_comparison} provides a comparison of some of the most notable datasets, highlighting their coverage in terms of number of speakers, synthesis method, and conversational structure.

\begin{table*}[ht]
  \centering
  \caption{Comparison of Existing Audio Deepfake Datasets}
  \label{tab:dataset_comparison}
  \begin{tabular}{@{}lcccccc@{}}
    \toprule
    \textbf{Dataset} & \textbf{\# Speakers} & \textbf{TTS} & \textbf{Conversational} & \textbf{Notes} \\
    \midrule
    ASVspoof 2019/2021 \cite{wang2020, delgado2021} & Single & \checkmark & $\boldsymbol{\times}$ & Standard benchmark for spoofed ASV; widely used \\
    \midrule
    WaveFake (2021) \cite{frank2021} & Single & \checkmark & $\boldsymbol{\times}$ & English/Japanese utterances generated by TTS \\
    \midrule
    In-the-Wild \cite{muller2022} & Single & \checkmark &  $\boldsymbol{\times}$ & Collected in uncontrolled/noisy environments \\
    \midrule
    ADD2022/2023 \cite{yi2022b,yi2023} & Single & \checkmark  & $\boldsymbol{\times}$ & Challenge datasets with partially fake tasks \\
    \midrule
    Fake-or-Real (FoR) \cite{reimao2019} & Single & \checkmark & $\boldsymbol{\times}$ & Binary classification (authentic vs fake) \\
    \midrule
    MADD \cite{qi2024} & Single & \checkmark & $\boldsymbol{\times}$ & Multilingual, multiple voices in the dataset but not in the same audio \\
    \midrule
    HAD \cite{yi2021} & Single & \checkmark  & $\boldsymbol{\times}$ & Partial manipulations (segment replacements) \\
    \midrule
    PartialSpoof \cite{zhang2023} & Single & \checkmark & $\boldsymbol{\times}$
 & Segment-level manipulations within utterances \\
 \midrule
 \textbf{MsCADD} & \textbf{Multi-speaker} & \checkmark & \checkmark& \textbf{Dataset involving multiple speakers in the same audio clips} \\
    \bottomrule
  \end{tabular}
\end{table*}

\section{Multi-Speaker Deepfake Taxonomy and Dataset}

A significant and unexplored gap in the field of audio deepfake detection is the lack of datasets specifically designed for multi-speaker environments. While existing benchmarks \cite{almutairi2022, khanjani2023} have contributed substantially to the progress of single-speaker deepfake detection, they fall short in capturing the conversational dynamics, interaction complexity, and contextual realism found in real-world multi-speaker audio.

To facilitate robust detection research and system development, we propose a structured framework for creating multi-speaker conversational audio deepfake dataset (MsCADD). This framework is grounded in a novel three-branch taxonomy that categorizes multi-speaker deepfakes by conversational context, speaker composition, and the scope of manipulation.

\subsection{Taxonomy of Multi-Speaker Conversational Audio Deepfake}

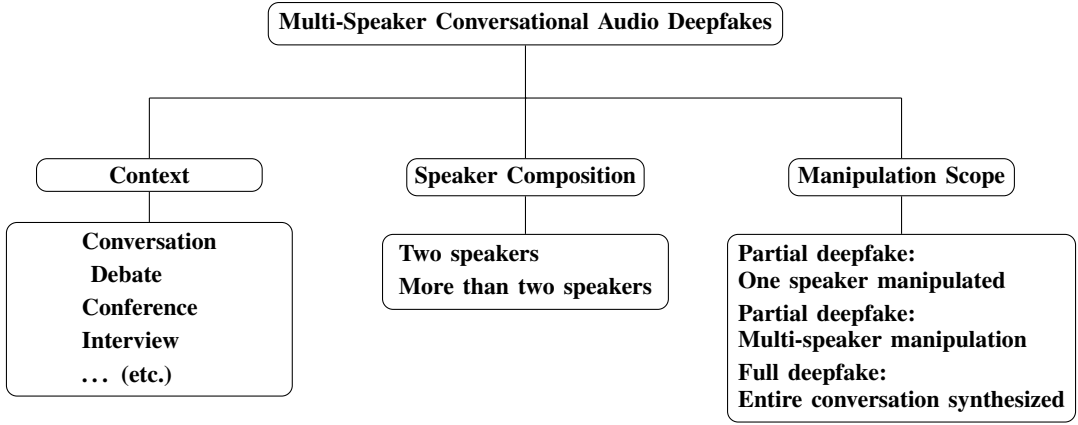
\begin{figure*}[t]
\centering
\begin{tikzpicture}[
  every node/.style={font=\small},
  titlebox/.style={draw, rounded corners, align=center, inner sep=3pt, minimum width=30mm},
  contentbox/.style={draw, rounded corners, align=left, inner sep=4pt, minimum width=38mm},
  rootbox/.style={draw, rounded corners, align=center, inner sep=4pt, minimum width=55mm},
  >={Latex[length=2mm]}
]

\node[rootbox] (root) {\textbf{Multi-Speaker Conversational Audio Deepfakes}};

\coordinate (split) at ($(root.south)+(0,-7mm)$);
\draw (root.south) -- (split);

\coordinate (cX) at ($(split)+(-50mm,0)$);
\coordinate (sX) at ($(split)+(0,0)$);
\coordinate (mX) at ($(split)+(50mm,0)$);
\draw (split) -- (cX);
\draw (split) -- (sX);
\draw (split) -- (mX);

\node[titlebox, below=8mm of cX] (ctxT) {\textbf{Context}};
\node[titlebox, below=8mm of sX] (spkT) {\textbf{Speaker Composition}};
\node[titlebox, below=8mm of mX] (manT) {\textbf{Manipulation Scope}};

\draw (cX) |- (ctxT.north);
\draw (sX) |- (spkT.north);
\draw (mX) |- (manT.north);

\node[contentbox, below=4mm of ctxT] (ctxC) {%
  \textbf{Conversation}\\[2.5pt]
 \textbf{ Debate}\\[2.5pt]
  \textbf{Conference}\\[2.5pt]
  \textbf{Interview}\\[2.5pt]
  \textbf{\ldots\ (etc.)}
};
\draw (ctxT.south) -- (ctxC.north);

\node[contentbox, below=5mm of spkT] (spkC) {%
  \textbf{Two speakers}\\[2.5pt]
  \textbf{More than two speakers}
};
\draw (spkT.south) -- (spkC.north);

\node[contentbox, below=5mm of manT] (manC) {%

  \textbf{Partial deepfake:} \\ \textbf{One speaker manipulated} \\[2.5pt]
  
  \textbf{Partial deepfake:}\\ \textbf{Multi-speaker manipulation} \\[2.5pt]
  
  \textbf{Full deepfake:} \\ \textbf{Entire conversation synthesized}
};
\draw (manT.south) -- (manC.north);

\path (ctxC.south) -- (spkC.south) coordinate[pos=0.5] (alignA);

\end{tikzpicture}
\caption{Taxonomy of Multi-Speaker Conversational Audio Deepfakes. This structure organizes deepfake scenarios based on interaction context, speaker composition, and manipulation scope.}
\label{fig:taxonomy}
\end{figure*}

The taxonomy in Figure~\ref{fig:taxonomy} classifies multi-speaker audio deepfake scenarios across three dimensions:

\begin{itemize}
    \item \textbf{Context:} The social or communicative setting in which the conversation occurs. Examples include casual conversations, debates, interviews, and conferences.
    \item \textbf{Speaker Composition:} The number of speakers present in the interaction. This is typically categorized into two speakers and three or more speakers.
    \item \textbf{Manipulation Scope:} The extent to which speech is manipulated. This includes:
    \begin{itemize}
    \item \textbf{Partial deepfake: One speaker manipulated:} Only one speaker's voice contains synthetic or manipulated segments, while all other speakers' voices remain fully authentic.
    
    \item \textbf{Partial deepfake: Multi-speaker manipulation:} Multiple speakers have parts of their speech altered or synthesized, but the entire conversation is not fully fake.
    
    \item \textbf{Full deepfake: Entire conversation synthesized:} Every speaker's voice and the complete conversation content are entirely synthesized without any authentic audio.
\end{itemize}

\end{itemize}

\subsection{Dataset Creation}
We initiate our study of multi-speaker conversational audio deepfakes with the manipulation scope—\textit{full deepfake}, where the entire conversation is synthetically generated. To support this investigation, we developed a new dataset (MsCADD) comprising both real conversations and fully synthetic conversations. The dataset consists of 2,830 audio clips in .wav format, organized into two classes: 1,148 real audios and 1,682 deepfake audios. The duration of the audios ranges from 10 seconds to 22 seconds. The content and source of MsCADD are described below.

\begin{enumerate}
    \item \textbf{Source Corpus}\\
    The conversational content was derived from the English Conversation Corpus \cite{li2022enhancing} introduced at IEEE International Conference on Acoustics, Speech and Signal Processing in 2022 (ICASSP 2022). This corpus collects 24 hours of speech from 66 publicly available conversational videos from YouTube. It contains natural multi-speaker dialogues that we used as text scripts for generating synthetic conversations. Each dialogue was segmented into speaker turns and assigned to synthetic voices to emulate realistic conversational flow.

    \item \textbf{Synthetic Audio Generation}\\
    We generated the synthetic conversations using two complementary TTS systems. VITS (Coqui TTS library) \cite{kim2021vits} was employed to produce diverse speaker voices, ensuring variability in timbre and accent across the dataset. We concatenated the speech segments with short pauses of approximately 0.5 seconds between turns to preserve the rhythm of real dialogues. In addition, we used NotebookLM’s SoundStorm-based generation (Google TTS) \cite{borsos2023soundstorm} to create more natural conversational exchanges. This choice allowed us to incorporate elements of spontaneity such as back-and-forth flow, laughter, and subtle conversational nuances that make the synthetic dialogues feel more human-like. Table~\ref{tab:audio_dataset} shows the audio dataset distribution. 

    \begin{table}[ht]
  \centering
  \caption{MsCADD Distribution}
  \label{tab:audio_dataset}
  \begin{tabular}{@{}lccc@{}}
  \toprule
    \textbf{Category}   & \textbf{Source}   & \textbf{\# Unique Voices} & \textbf{\# Audios} \\
    \midrule
    Fake & VITS              & 5   & 573  \\
             Fake           & SoundStorm        & 2   & 1109 \\
             \midrule
                    \textbf{Total Fake}    & --- & 7   & 1682 \\
            \midrule
    Real                & Human recordings  & 5+  & 1148 \\ 
    \midrule
    \textbf{Grand Total}& ---               & 12+ & 2830 \\
    \bottomrule
  \end{tabular}
\end{table}

    \item \textbf{Speaker Composition}\\
    Each conversation clip features two speakers, with gender combinations randomized:
    \begin{itemize}
        \item Male--Female
        \item Female--Female
        \item Male--Male
    \end{itemize}
    Voices were randomly sampled from the TTS speaker pools, ensuring that no single voice dominated the dataset. Across the dataset, multiple unique synthetic voices were used, and real speakers were similarly diverse to capture accents and variations. Table~\ref{tab:breakdown_speaker} shows the breakdown of speaker conposition for the fake audios.

    \begin{table}[ht]
  \centering
  \caption{Breakdown of Speaker Composition}
  \label{tab:breakdown_speaker}
  \begin{tabular}{@{}llr@{}}
    \toprule
    \textbf{Source}   & \textbf{Speaker Composition}
   & \textbf{\# Audios} \\
    \midrule
    VITS & Male--Female           & 172  \\
    VITS    & Female--Female              & 166  \\
        VITS & Male--Male                 & 235  \\
    \midrule
    SoundStorm        & Male \& Female           & 1109 \\
    \midrule
    
    \textbf{Total} & --- & 1682 \\
    \bottomrule
  \end{tabular}
\end{table}\

\item \textbf{Real Audio Collection}\\
The real subset of 1,148 audio clips consists of genuine human-recorded conversations with similar two-speaker structures and varied content. The audios comes from the English Conversation Corpus \cite{li2022enhancing}. These serve as the real audio ground truth for benchmarking.
\end{enumerate}

The dataset has been released on our GitHub. The GitHub repository includes data samples, the full dataset and codes are available on request. This ensures that other researchers can reproduce our results and extend the benchmark for future studies.

\section{Experimental Results}

In this section, we evaluate the performance of our dataset using neural baseline models that have been widely applied in audio deepfake detection and spoofing challenges. We discuss the results with (a) LFCC--LCNN (b) RawNet2 and (c) pretrained Wav2Vec 2.0. We split the dataset into 80\% training and 20\% testing, stratified by class. We report accuracy (percentage of correctly classified instances), true positive rate (TPR), True negative rate (TNR) and F1 score for fake class.
The purpose is to establish reference points against which future methods for multi-speaker deepfake detection can be compared.

\subsection{Linear Frequency Ceptral Coefficients (LFCC)--Light Convolutional Neural Networks (LCNN) Baseline}
Linear Frequency Ceptral Coefficients (LFCC) \cite{yamagishi2021asvspoof} is the input of the Light Convolutional Neural Networks (LCNN) \cite{sahidullah2015comparison} as our baseline model. The LFCC-LCNN model was adapted from the official ASVspoof 2021 baseline \cite{yamagishi2021asvspoof}. This model is one of the well performing baselines in the challenge \cite{delgado2021} for Deepfake and Logical Access tasks\cite{khanjani2023b}. We trained and evaluated this baseline on our dataset. The model achieved 0.65 and 0.62 as the f1 score and accuracy on the testing data respectively. To determine the decision threshold based on the model’s output scores, we empirically evaluated a range of thresholds and compared their corresponding performance metrics. Based on this analysis, a threshold of –15.82 was selected. The true positive rate (TPR) for fake speech was 92.80\%, while the true negative rate (TNR) for real speech was 53.40\%.

\subsection{RawNet2 Baseline}
RawNet2 \cite{tak2021rawnet2} is an end-to-end convolutional neural network originally proposed for the ASVspoof challenge \cite{delgado2021}. Unlike feature-based models, RawNet2 directly ingests raw waveforms, with early convolutional layers acting as learnable band-pass filters and gated recurrent layers modeling temporal dependencies. We adapted the official ASVspoof 2021 baseline \cite{yamagishi2021asvspoof} implementation of RawNet2 to our dataset. On the test set, the model achieved an F1 score of 0.88 and an accuracy of 0.90. The true positive rate (TPR) for fake speech was 84.27\%, while the true negative rate (TNR) for real speech was 97.39\%.

\subsection{Wav2Vec 2.0}
Wav2Vec 2.0 \cite{baevski2020wav2vec} is a self-supervised speech representation model pretrained on large-scale audio data. We used a publicly available checkpoint fine-tuned for deepfake detection \cite{alexandreacff_wav2vec2_fake} as our baseline model. The model outputs per-segment fake probabilities, which were aggregated across each clip. The model achieved an F1 score of 0.89 and an accuracy of 0.89. The true positive rate (TPR) for fake speech was 82.5\%, while the true negative rate (TNR) for real speech was 97.8\%.

\subsection{Overall Findings}

The results of the three baseline neural models on the multi-speaker conversational audio deepfake dataset (MsCADD) are summarized in Table~\ref{tab:baseline_results}. Overall, we observe that while all models demonstrate the ability to distinguish real from fully synthetic conversations, their performance characteristics differ in meaningful ways.

The LFCC-LCNN baseline, adapted from the ASVspoof 2021 challenge, achieved moderate performance. Although it reached a relatively high true positive rate (TPR) for fake speech, its true negative rate (TNR) on real conversations was comparatively low. This indicates that LFCC-LCNN tends to over-predict “fake” when applied to conversational settings, struggling to generalize to genuine multi-speaker speech.

In contrast, RawNet2 achieved good and balanced performance. Its TPR for fake speech and TNR for real speech shows that RawNet2 effectively captured discriminative cues in both real and synthetic samples. Its slightly lower TPR on fake speech shows that there is room for improvement in detecting synthetic voices.

Finally, Wav2Vec 2.0 performance highlights the strength of large-scale self-supervised pretraining, which provides robust generalization across unseen conversational contexts. The model’s lower TPR on fake speech indicates that detecting synthetic voices remains an area for improvement.

The findings indicate that classic convolutional baselines such as LFCC-LCNN underperform in multi-speaker conversational conditions, whereas modern end-to-end models (RawNet2) and pretrained self-supervised models (Wav2Vec 2.0) exhibit stronger robustness. However, improvements are still required for reliably detecting synthetic voices in diverse conversational dynamics. This highlights the need for detection strategies that leverage conversational structure to improve performance in multi-speaker deepfake scenarios. We did not conduct detailed breakdowns such as breakdown by TTS system or speaker gender composition, but such analyses may reveal model-specific vulnerabilities and will be explored in future work.

\begin{table}[ht]
  \centering
  \caption{Performance of Neural Baseline Models on Multi-Speaker Deepfake Dataset}
  \label{tab:baseline_results}
  \begin{tabular}{@{}lcccc@{}}
    \toprule
    \textbf{Model} & \textbf{F1 Score} & \textbf{Accuracy} & \textbf{TPR (Fake)} & \textbf{TNR (Real)} \\
    \midrule
    LFCC-LCNN    & 0.65 & 0.62 & 92.80\% & 53.40\% \\
    RawNet2      & 0.88 & 0.90 & 84.27\% & 97.39\% \\
    Wav2Vec 2.0  & 0.89 & 0.89 & 82.50\% & 97.80\% \\
    \bottomrule
  \end{tabular}
\end{table}

\section{Conclusion and Future Work}

In this paper, we introduced the unexplored problem of detecting multi-speaker conversational audio deepfakes. We introduced a taxonomy of manipulation scopes, ranging from partial to full conversation deepfakes, and presented a new dataset of 2,830 multi-speaker conversations, comprising both real and synthetic dialogues. Using this dataset, we benchmarked three widely used neural baselines (LFCC-LCNN, RawNet2, and Wav2Vec 2.0), showing that modern end-to-end and self-supervised models outperform traditional convolutional baselines but still face challenges in reliably detecting synthetic voices under varied conversational dynamics.

Our findings emphasize two key takeaways. First, classic baselines such as LFCC-LCNN degrade significantly in multi-speaker conditions, highlighting the limitations of models designed primarily for single-speaker detection tasks. Second, although RawNet2 and Wav2Vec 2.0 achieve good results, their lower true positive rate (TPR) on fake speech indicates that synthetic voices remain difficult to detect in realistic, and dynamic dialogues.

This pilot study created a foundation for discussing these new types of deepfakes. For our future work, this study opens several directions for further research:

\begin{itemize}
    \item We plan to expand our benchmark comparisons to include recent transformer-based deepfake detection models such as those based on Audio Spectrogram Transformer (AST) \cite{Le2024ContinuousLearning, Gong2025DeepfakeVoice, Tahaoglu2025NeXtTDNN} or Conformer architectures \cite{Shin2023HMConformer} and explore multi-modal approaches that combine acoustic and linguistic features.
    \item \textbf{Partial Manipulation Scope:} Our dataset focused on the full deepfake scope, but real-world threats are more likely to involve partial manipulations, where only one or a subset of speakers are altered. Future work will extend the dataset to include these scenarios, enabling research into detecting manipulated segments within otherwise genuine conversations.
    \item \textbf{Conversational Context Modeling:} We will continue to investigate models that leverage dialogue structure, turn-taking patterns, and conversational cues (e.g., interruptions, laughter, pauses) to better capture the authenticity of multi-speaker exchanges.
    \item \textbf{A New Detection Model for Multi-Speaker Deepfakes:} As a next step, we will design and evaluate a model tailored specifically for multi-speaker detection, with the goal of better capturing speaker interactions and improving sensitivity to synthetic voices in complex conversational dynamics.
\end{itemize}

Finally, the MsCADD dataset sample has been made publicly available, enabling the community to build upon our work and accelerate progress in multi-speaker deepfake detection. The materials are made available for research and should be used responsibly.

\section{Acknowledgments}
Authors would like to acknowledge support from the National Science Foundation Award \#2346473.


{
\bibliographystyle{unsrt}
\bibliography{bibl.bib}

@misc{smith2021,
  author    = {B. Smith},
  title     = {Goldman Sachs, Ozy Media and a \$40 Million Conference Call Gone Wrong},
  year      = {2021},
  url       = {https://www.nytimes.com/2021/09/26/business/media/ozy-mediagoldman-sachs.html},
  note      = {[Online; accessed January 11, 2023]},
  journal   = {The New York Times}
}

@misc{harwell2021,
  author    = {D. Harwell},
  title     = {Remember the ‘Deepfake Cheerleader Mom’? Prosecutors Now Admit They Can’t Prove Fake-Video Claims},
  year      = {2021},
  note      = {March 14, 2021}
}

@misc{stupp2022,
  author    = {C. Stupp},
  title     = {Fraudsters Used AI to Mimic CEO’s Voice in Unusual Cybercrime Case},
  year      = {2022},
  url       = {https://www.wsj.com/articles/fraudsters-use-ai-to-mimic-ceos-voice-in-unusual-cybercrime-case-11567157402},
  note      = {[Online; accessed January 29, 2022]},
  journal   = {The Wall Street Journal}
}

@inproceedings{wang2018,
  author    = {Y. Wang and D. Stanton and Y. Zhang and et al.},
  title     = {Style Tokens: Unsupervised Style Modeling, Control and Transfer in End-to-End Speech Synthesis},
  booktitle = {Proc. of ICML},
  year      = {2018}
}

@article{sisman2020,
  author    = {B. Sisman and J. Yamagishi and S. King and H. Li},
  title     = {An Overview of Voice Conversion and Its Challenges: From Statistical Modeling to Deep Learning},
  journal   = {IEEE/ACM Transactions on Audio, Speech, and Language Processing},
  volume    = {29},
  pages     = {132--157},
  year      = {2020}
}

@inproceedings{shen2018,
  author    = {J. Shen and R. Pang and Ron J. Weiss and et al.},
  title     = {Natural TTS Synthesis by Conditioning Wavenet on Mel Spectrogram Predictions},
  booktitle = {Proc. of ICASSP},
  year      = {2018}
}

@article{jin2017,
  author    = {Z. Jin and G. J. Mysore and S. DiVerdi and J. Lu and A. Finkelstein},
  title     = {VoCo: Text-Based Insertion and Replacement in Audio Narration},
  journal   = {ACM Transactions on Graphics},
  volume    = {36},
  number    = {4},
  article   = {96},
  year      = {2017},
  pages     = {13},
  url       = {https://doi.org/10.1145/3072959.3073702}
}

@inproceedings{kim2021,
  author    = {J. Kim and J. Kong and J. Son},
  title     = {Conditional Variational Autoencoder with Adversarial Learning for End-to-End Text-to-Speech},
  booktitle = {International Conference on Machine Learning},
  year      = {2021},
  pages     = {5530--5540},
  publisher = {PMLR}
}

@inproceedings{wang2017,
  author    = {Y. Wang and R. J. Skerry-Ryan and D. Stanton and Y. Wu and R. A. Saurous},
  title     = {Tacotron: Towards End-to-End Speech Synthesis},
  booktitle = {Proc. of INTERSPEECH},
  year      = {2017}
}

@inproceedings{wang2021,
  author    = {T. Wang and R. Fu and J. Yi and J. Tao and S. Wang},
  title     = {Prosody and Voice Factorization for Few-Shot Speaker Adaptation in the Challenge M2VOC 2021},
  booktitle = {Proc. of ICASSP},
  year      = {2021}
}

@article{tan2021,
  author    = {X. Tan and T. Qin and F. Soong and T.-Y. Liu},
  title     = {A Survey on Neural Speech Synthesis},
  journal   = {arXiv preprint arXiv:2106.15561},
  year      = {2021}
}

@inproceedings{popov2021,
  author    = {V. Popov and I. Vovk and V. Gogoryan and T. Sadekova and M. Kudinov},
  title     = {Grad-TTS: A Diffusion Probabilistic Model for Text-to-Speech},
  booktitle = {International Conference on Machine Learning},
  year      = {2021},
  pages     = {8599--8608},
  publisher = {PMLR}
}

@article{khanjani2023,
  author    = {Z. Khanjani and G. Watson and V. P. Janeja},
  title     = {Audio Deepfakes: A Survey},
  journal   = {Frontiers in Big Data},
  volume    = {5},
  article   = {1001063},
  year      = {2023},
  doi       = {10.3389/fdata.2022.1001063}
}

@article{almutairi2022,
  author    = {Z. Almutairi and H. Elgibreen},
  title     = {A Review of Modern Audio Deepfake Detection Methods: Challenges and Future Directions},
  journal   = {Algorithms},
  volume    = {15},
  number    = {5},
  article   = {155},
  year      = {2022},
  doi       = {10.3390/a15050155}
}

@article{todisco2019,
  author    = {M. Todisco and X. Wang and V. Vestman and M. Sahidullah and H. Delgado and A. Nautsch and J. Yamagishi and N. Evans and T. Kinnunen and K. A. Lee},
  title     = {ASVspoof 2019: Future Horizons in Spoofed and Fake Audio Detection},
  journal   = {arXiv},
  year      = {2019},
  arxiv     = {1904.05441}
}

@article{todisco2019b,
  author    = {M. Todisco and X. Wang and V. Vestman and M. Sahidullah and H. Delgado and A. Nautsch and J. Yamagishi and N. Evans and T. Kinnunen and K. A. Lee},
  title     = {ASVspoof 2019: Future Horizons in Spoofed and Fake Audio Detection},
  journal   = {arXiv},
  year      = {2019},
  arxiv     = {1904.05441}
}

@article{liu2021,
  author    = {T. Liu and D. Yan and R. Wang and N. Yan and G. Chen},
  title     = {Identification of Fake Stereo Audio Using SVM and CNN},
  journal   = {Information},
  volume    = {12},
  article   = {263},
  year      = {2021},
  doi       = {10.3390/info12070263}
}

@inproceedings{khanjani2023b,
  author    = {Z. Khanjani and L. Davis and A. Tuz and K. Nwosu and C. Mallinson and V. P. Janeja},
  title     = {Learning to Listen and Listening to Learn: Spoofed Audio Detection through Linguistic Data Augmentation},
  booktitle = {IEEE International Conference on Intelligence and Security Informatics (ISI)},
  year      = {2023},
  pages     = {979--8--3503--3773-0}
}

@misc{wired-slovakia-deepfakes,
  title        = {Slovakia’s Election Deepfakes Show AI Is a Danger to Democracy},
  author       = {Morgan Meaker},
  year         = {2023},
  month        = oct,
  day          = {3},
  url          = {https://www.wired.com/story/slovakias-election-deepfakes-show-ai-is-a-danger-to-democracy/},
  note         = {Accessed 2025-08-25},
}

@misc{cnn-election-deepfake-threats,
  title        = {Election deepfake threats: CNN coverage},
  author       = {CNN},
  year         = {2024},
  month        = feb,
  day          = {1},
  url          = {https://www.cnn.com/2024/02/01/politics/election-deepfake-threats-invs/index.html},
  note         = {Accessed 2025-08-25},
}

@article{khochare2021deep,
  author    = {Khochare, J. and Joshi, C. and Yenarkar, B. and Suratkar, S. and Kazi, F.},
  title     = {A Deep Learning Framework for Audio Deepfake Detection},
  journal   = {Arabian Journal for Science and Engineering},
  year      = {2021},
  pages     = {1--12},
  publisher = {Springer}
}

@article{oord2016,
  author    = {A. Oord and S. v. d. Dieleman and H. Zen and K. Simonyan and O. Vinyals and et al.},
  title     = {WaveNet: A Generative Model for Raw Audio},
  journal   = {arXiv preprint arXiv:1609.03499 [cs]},
  year      = {2016},
  doi       = {10.48550/arXiv.1609.03499}
}

@inproceedings{li2022enhancing,
  author    = {Li, Jingbei and Meng, Yi and Li, Chenyi and Wu, Zhiyong and Meng, Helen and Weng, Chao and Su, Dan},
  title     = {Enhancing Speaking Styles in Conversational Text-to-Speech Synthesis with Graph-Based Multi-Modal Context Modeling},
  booktitle = {2022 IEEE International Conference on Acoustics, Speech and Signal Processing (ICASSP)},
  pages     = {7917--7921},
  year      = {2022},
  organization = {IEEE},
  doi       = {10.1109/ICASSP43922.2022.9747837}
}

@inproceedings{kim2021vits,
  title     = {Conditional Variational Autoencoder with Adversarial Learning for End-to-End Text-to-Speech},
  author    = {Kim, Jaehyeon and Kong, Jungil and Son, Juhee},
  booktitle = {Proceedings of the 38th International Conference on Machine Learning (ICML)},
  year      = {2021},
  volume    = {139},
  pages     = {5530--5540},
  publisher = {PMLR},
  url       = {https://arxiv.org/abs/2106.06103}
}

@article{borsos2023soundstorm,
  title   = {SoundStorm: Efficient Parallel Audio Generation},
  author  = {Borsos, Zal{\'a}n and Sharifi, Matt and Vincent, Damien and Kharitonov, Eugene and Zeghidour, Neil and Tagliasacchi, Marco},
  journal = {arXiv preprint arXiv:2305.09636},
  year    = {2023},
  url     = {https://arxiv.org/abs/2305.09636}
}

@article{wang2020,
  author    = {X. Wang and J. Yamagishi and M. Todisco and H. Delgado and A. Nautsch and N. Evans and \textit{et al.}},
  title     = {ASVspoof 2019: A Large-Scale Public Database of Synthesized, Converted and Replayed Speech},
  journal   = {Comput. Speech Lang.},
  volume    = {64},
  article   = {101114},
  year      = {2020},
  doi       = {10.1016/j.csl.2020.101114}
}

@misc{delgado2021,
  author    = {H. Delgado and N. Evans and T. Kinnunen and K. A. Lee and X. Liu and A. Nautsch \textit{et al.}},
  title     = {ASVspoof 2021: Automatic Speaker Verification Spoofing and Countermeasures Challenge, v0.3},
  year      = {2021},
  url       = {https://www.asvspoof.org/asvspoof2021/asvspoof2021_evaluation_plan.pdf},
  note      = {[Online; accessed February 2023]}
}

@article{frank2021,
  author    = {J. Frank and L. Schönherr},
  title     = {WaveFake: A Data Set to Facilitate Audio Deepfake Detection},
  journal   = {arXiv [Preprint]},
  year      = {2021},
  arxiv     = {2111.02813}
}

@article{muller2022,
  author    = {N. M. Müller and P. Czempin and F. Dieckmann and A. Froghyar and K. Böttinger},
  title     = {Does Audio Deepfake Detection Generalize?},
  journal   = {Interspeech},
  year      = {2022}
}

@inproceedings{yi2022b,
  author    = {J. Yi and R. Fu and J. Tao and S. Nie and H. Ma and C. Wang and T. Wang and Z. Tian and Y. Bai and C. Fan and et al.},
  title     = {ADD 2022: The First Audio Deep Synthesis Detection Challenge},
  booktitle = {ICASSP 2022--2022 IEEE International Conference on Acoustics, Speech and Signal Processing (ICASSP)},
  year      = {2022},
  pages     = {9216--9220},
  publisher = {IEEE}
}

@misc{yi2023,
  author    = {J. Yi and J. Tao and R. Fu and X. Yan and C. Wang and T. Wang and C. Zhang and X. Zhang and Y. Zhao and Y. Ren and et al.},
  title     = {ADD 2023: The Second Audio Deepfake Detection Challenge},
  year      = {2023},
  note      = {IJCAI 2023 Workshop on Deepfake Audio Detection and Analysis (DADA 2023), August 19, 2023, Macao, S.A.R.}
}

@inproceedings{reimao2019,
  author    = {R. Reimao and V. Tzerpos},
  title     = {For: A Dataset for Synthetic Speech Detection},
  booktitle = {2019 International Conference on Speech Technology and Human Computer Dialogue (SpeD)},
  year      = {2019},
  pages     = {1--10},
  publisher = {IEEE},
  doi       = {10.1109/SPED.2019.8906599}
}

@inproceedings{qi2024,
  author    = {X. Qi and H. Gu and J. Yi and J. Tao and Y. Ren and J. He and S. Zeng},
  title     = {MADD: A Multi-lingual Multi-speaker Audio Deepfake Detection Dataset},
  booktitle = {2024 IEEE 14th International Symposium on Chinese Spoken Language Processing (ISCSLP)},
  year      = {2024},
  pages     = {1--6},
  publisher = {IEEE},
  doi       = {10.1109/ISCSLP63861.2024.10800535}
}

@article{yi2021,
  author    = {J. Yi and Y. Bai and J. Tao and H. Ma and Z. Tian and C. Wang and T. Wang and R. Fu},
  title     = {Half-Truth: A Partially Fake Audio Detection Dataset},
  journal   = {arXiv preprint arXiv:2104.03617},
  year      = {2021}
}

@article{zhang2023,
  author    = {L. Zhang and X. Wang and E. Cooper and N. Evans and J. Yamagishi},
  title     = {The PartialSpoof Database and Countermeasures for the Detection of Short Fake Speech Segments Embedded in an Utterance},
  journal   = {IEEE/ACM Transactions on Audio, Speech, and Language Processing},
  volume    = {31},
  year      = {2023},
  doi       = {10.1109/TASLP.2022.3233236}
}

@inproceedings{chen2017resnet,
  author    = {Chen, Z. and Xie, Z. and Zhang, W. and Xu, X.},
  title     = {ResNet and Model Fusion for Automatic Spoofing Detection},
  booktitle = {Interspeech 2017},
  pages     = {102--106},
  year      = {2017},
  organization = {ISCA}
}

@inproceedings{chen2020generalization,
  author    = {Chen, T. and Kumar, A. and Nagarsheth, P. and Sivaraman, G. and Khoury, E.},
  title     = {Generalization of Audio Deepfake Detection},
  booktitle = {Odyssey 2020 The Speaker and Language Recognition Workshop},
  pages     = {132--137},
  year      = {2020},
  organization = {ISCA}
}

@article{yamagishi2021asvspoof,
  title   = {ASVspoof 2021: Accelerating progress in spoofed and deepfake speech detection},
  author  = {Yamagishi, Junichi and Wang, Xin and Todisco, Massimiliano and Sahidullah, Md and Patino, Jose and Nautsch, Andreas and Liu, Xin and Lee, Kong Aik and Kinnunen, Tomi and Evans, Nicholas and others},
  journal = {arXiv preprint arXiv:2109.00537},
  year    = {2021},
  url     = {https://arxiv.org/abs/2109.00537}
}

@inproceedings{sahidullah2015comparison,
  author    = {Sahidullah, Md and Kinnunen, Tomi and Hanil{\c{c}}i, Cemal},
  title     = {A Comparison of Features for Synthetic Speech Detection},
  booktitle = {Proc. Interspeech},
  year      = {2015},
  organization = {ISCA}
}

@article{tak2021rawnet2,
  title   = {End-to-End Anti-Spoofing with RawNet2},
  author  = {Tak, Hemlata and Patino, Jose and Todisco, Massimiliano and Nautsch, Andreas and Evans, Nicholas and Larcher, Anthony},
  journal = {arXiv preprint arXiv:2011.01108},
  year    = {2021},
  url     = {https://arxiv.org/abs/2011.01108}
}

@article{baevski2020wav2vec,
  title   = {wav2vec 2.0: A Framework for Self-Supervised Learning of Speech Representations},
  author  = {Baevski, Alexei and Zhou, Henry and Mohamed, Abdelrahman and Auli, Michael},
  journal = {arXiv preprint arXiv:2006.11477},
  year    = {2020},
  url     = {https://arxiv.org/abs/2006.11477}
}

@misc{alexandreacff_wav2vec2_fake,
  author       = {Alexandreacff},
  title        = {wav2vec2-base-ft-fake-detection},
  howpublished = {\url{https://huggingface.co/alexandreacff/wav2vec2-base-ft-fake-detection}},
  note         = {Accessed: 2025-08-26}
}

@article{Le2024ContinuousLearning,
  author       = {Le, Tuan Duy Nguyen and Teh, Kah Kuan and Tran, Huy Dat},
  title        = {Continuous Learning of Transformer-based Audio Deepfake Detection},
  journal      = {arXiv preprint arXiv:2409.05924},
  year         = {2024},
  url          = {https://arxiv.org/abs/2409.05924},
}

@article{Gong2025DeepfakeVoice,
  author       = {Gong, Liang Yu and Li, Xue Jun},
  title        = {Deepfake Voice Detection: An Approach Using End-to-End Transformer with Acoustic Feature Fusion by Cross-Attention},
  journal      = {Electronics},
  volume       = {14},
  number       = {10},
  pages        = {2040},
  year         = {2025},
  doi          = {10.3390/electronics14102040},
  url          = {https://doi.org/10.3390/electronics14102040},
  publisher    = {MDPI},
  note         = {Open Access under CC BY 4.0},
}

@article{Tahaoglu2025NeXtTDNN,
  author       = {Tahaoglu, Gul},
  title        = {Robust DeepFake Audio Detection via an Improved NeXt-TDNN with Multi-Fused Self-Supervised Learning Features},
  journal      = {Applied Sciences},
  volume       = {15},
  number       = {17},
  pages        = {9685},
  year         = {2025},
  doi          = {10.3390/app15179685},
  url          = {https://doi.org/10.3390/app15179685},
  publisher    = {MDPI},
  note         = {Open Access under CC BY 4.0},
}

@article{Shin2023HMConformer,
  author       = {Shin, Hyun-seo and Heo, Jungwoo and Kim, Ju-ho and Lim, Chan-yeong and Kim, Wonbin and Yu, Ha-Jin},
  title        = {HM-CONFORMER: A Conformer-based Audio Deepfake Detection System with Hierarchical Pooling and Multi-Level Classification Token Aggregation Methods},
  journal      = {arXiv preprint arXiv:2309.08208},
  year         = {2023},
  url          = {https://arxiv.org/abs/2309.08208},
}
}
\nocite{*}
\end{document}